\documentclass[aps,prb, superscriptaddress,twocolumn,floatfix,preprintnumbers,showpacs]{revtex4}
\usepackage{graphicx,amsmath,amssymb,bm,hyperref,color,engord,times, epstopdf}
\definecolor{blue}{rgb}{0.0,0.0,1.0}
\definecolor{black}{rgb}{0.0,0.0,0.0}
\definecolor{red}{rgb}{1.0,0.0,0.0}
\hypersetup{colorlinks=true,breaklinks, linkcolor=blue,urlcolor=blue,citecolor=red}
\begin{document}

\title{Measuring spin and charge correlations via tunneling-current conductance fluctuations}
\author{Kelly R. Patton}
\email[\hspace{-1.4mm}]{kpatton@physnet.uni-hamburg.de}
\author{Hartmut Hafermann}
\author{Sergej Brener}
\author{Alexander I. Lichtenstein}
\affiliation{I. Institut f\"ur Theoretische Physik Universit\"at Hamburg, Hamburg 20355, Germany }
\author{Mikhail I. Katsnelson}
\affiliation{Institute for Molecules and Materials, Radboud
University Nijmegen, Nijmegen 6525AJ, The Netherlands}
\date{\today}


\begin{abstract} Scanning tunneling miscoscopy is one of the most powerful
spectroscopic tools for single-electron excitations. We show that
the conductance fluctuations, or noise in the conductance, of a
tunneling current into an interacting electron system is dominated
by density-density and spin-spin correlations. This allows one to probe two-particle
properties (susceptibilities) and collective excitations by standard
experimental tunneling methods. We demonstrate this
theoretically, using a novel  many-body calculation
for the multi-center Kondo problem, including both direct and
indirect exchange between magnetic atoms. An example of the
two-particle correlations around a single magnetic adatom in the
Kondo regime, as would be viewed by a scanning tunneling microscope,
is given.  The spatial dependance of the charge and spin
correlations, including the formation of the Kondo cloud in the
spin sector, are shown.
\end{abstract}
\maketitle
Noise spectroscopy, typically current noise, has become an
exceedingly useful tool in the study of electronic systems.  The
intrinsic noise, i.e.~the fluctuations of a signal due to inherent uncertainties,
generated by an electronic system is not a simple set of
random uncorrelated events but contains  fundamental information
about electron-electron correlations, that is not seen in the (averaged)
signal itself. Perhaps, the most well-known demonstration  of this was the experimental verification of
fractionally charged quasi-particles in the quantum Hall regime,
by shot-noise measurements
\cite{KanePRL94a,SaminadayarPRL97}.
The study of current-current  correlations, such as shot-noise or
Johnson-Nyquist (thermal) noise, has a long history and has been
a prevalent topic of both experimental and theoretical
investigations, along with work in related areas,
such as conductance fluctuations of mesoscopic wires, mostly
by studying the effects of disorder
\cite{LeePRL85,vanOudenaardenPRL97,vonOppenPRL97}.  Here, we show that in the weak-local-tunneling limit the
conductance fluctuations of an unpolarized (spin) current into an
interacting system is dominated by density-density correlations or
for a spin-polarized current by the spin-spin correlations.  This
allows one to extract two-particle characteristics from tunneling
experiments along with single-particle quantities, such as the
density of states or magnetization. Macroscopic two-particle
properties, like the compressibility or magnetic susceptibility,
are easily measurable. However, few if any techniques exists to
measure these quantities at a local microscopic scale, with
spatial and energy resolution. For instance, almost fifty-years
after the discovery and theoretical explanation of the Kondo
effect, the spatially localized spin correlations around  a
magnetic impurity---the Kondo cloud---has never been
experimentally observed.

The atomic spatial resolution of a scanning tunneling microscope
(STM) makes it a natural choice to study local
fluctuations on the microscopic scale. This combination of an STM and noise spectroscopy has already been used to develop the field of electron spin 
resonance scanning tunneling microscopy (ESR-STM)\cite{ManassenPRL89}.   With a similar experimental setup in mind we will limit  
ourselves to an STM system, along with making the common approximations, such
as weak local tunneling and where the STM is assumed to be a weakly
or non-correlated Fermi liquid with a featureless single-particle
density of states. Although, the work presented here is with
reference to an STM, the results are not limited to such systems and are valid anywhere such approximations can be made.

The total Hamiltonian is taken to be
\begin{equation*}
H=H^{}_{\rm STM}+H^{}_{\rm sub}+H^{}_{\rm tun},
\end{equation*}
where $H^{}_{\rm sub}$ is the general interacting Hamiltonian of
the substrate and $H^{}_{\rm STM}=\sum_{{\bf
k}\sigma}(\epsilon^{}_{{\bf k}\sigma}-\mu-eV)a^{\dagger}_{{\bf
k}\sigma}a^{}_{{\bf k}\sigma}$ is the Hamiltonian of the STM. As usual, we assume that the STM is a
noninteracting Fermi system with an energy independent density of states near the Fermi energy $\epsilon^{}_{\rm F}=\mu$.  This is an optimal condition for spectroscopic aims. The chemical
potential of the STM is displaced by $eV$, where the charge of the
electron is $-e$ and $V$ is the applied voltage. The tunneling is
determined by
 $ H^{}_{\rm tun}=\sum_{{\bf k},{\bf k}',\sigma}\big[{\rm T}a^{\dagger}_{{\bf k}\sigma}b^{}_{{\bf k}'\sigma}+{\rm H.c.}\big]$,
with tunneling amplitude ${\rm T}$, which is assumed to be
independent of momenta ${\bf k}, {\bf k}'$. The
operators $b^{\dagger}_{{\bf k}'\sigma}$ and $b^{}_{{\bf
k}'\sigma} $ are the mode creation and annihilation operators for
the substrate. The current operator is defined as
$\hat{I}=-e\partial^{}_{t}{\hat N}_{\rm STM}$, where ${\hat
N}_{\rm STM}=\sum_{{\bf k}\sigma}a^{\dagger}_{{\bf
k}\sigma}a^{}_{{\bf k}\sigma}$ is the particle number operator for
the STM. Assuming $\big[{\hat N}_{\rm STM},H^{}_{\rm sub}\big]=0$,
by Heisenberg's equation of motion, $\hat{I}=ie\big[{\hat N}_{\rm
STM},H^{}_{\rm tun}\big]$ (with $\hbar =1$).  Evaluating the
commutator using $\big\{a^{}_{{\bf k}\sigma},b^{\dagger}_{{\bf
k}'\sigma'}\big\}=0 $ leads to, within the tunneling Hamiltonian
formalism, the common expression for the current operator\cite{Mahanbook}
\begin{equation}
\label{full current operator}
{\hat I}=ie \sum_{\sigma}\sum_{{\bf k},{\bf k}'}\big[{\rm T}a^{\dagger}_{{\bf k}\sigma}b^{}_{{\bf k}'\sigma}-{\rm H.c.}\big].
\end{equation}
To obtain the experimentally measured current, one needs to obtain
the non-equilibrium expectation value of (\ref{full current
operator}).  We do this within the linear response (LR) regime,
treating the tunneling as the perturbation and assuming the STM
and substrate are separately in thermodynamic equilibrium. If the
system is decoupled in the infinite past, $t=-\infty$, the current
operator within LR is given as
\begin{equation}
\label{LR current operator}
\hat{I}^{}_{\rm LR}(t)=\hat{I}+i\int\limits_{-\infty}^{t}dt'\, \big[H^{}_{\rm tun}(t'),{\hat I}(t)\big],
\end{equation}
where $\hat{O}(t)=e^{iH^{}_{0}t}\hat{O}e^{-iH^{}_{0}t}$ and
$H^{}_{0}=H^{}_{\rm STM}+H^{}_{\rm sub}$. The expectation value of
(\ref{LR current operator})   with respect to $H_{0}$,
$\big<\hat{I}^{}_{\rm LR}\big>_{H_{0}}={\rm Tr}\, \hat{I}^{}_{\rm
LR}e^{-\beta H^{}_{0}}/{\rm Tr}\, e^{-\beta H^{}_{0}}$, gives the
current to leading order in the tunneling amplitude, $T$.
Therefore, the linear conductance, in this approximation, is given
by $G^{}_{\rm LR}=\partial^{}_{V}\big<\hat{I}^{}_{\rm
LR}\big>_{H_{0}}$. Assuming that the expectation value and derivative
commute, a conductance operator can also be defined by
$\hat{G}^{}_{\rm LR}=\partial^{}_{V}\hat{I}^{}_{\rm LR}$ (see supplemental material for details).  
With an operator expression for the conductance, one can obtain the  fluctuations or specifically the spectral
density  of the conductance, defined as
\begin{align}
\label{noise def}
S({\bf r},\omega)&=\frac{1}{2}\int dt\, e^{i\omega t}\big<\big\{\delta \hat{G}_{\rm LR}({\bf r},t),\delta \hat{G}_{\rm LR}({\bf r},0)\big\}\big>_{H_{0}},
\end{align}
where $\delta \hat{G}_{\rm LR}=\hat{G}_{\rm LR}-\big<\hat{G}_{\rm LR}\big>$. The low-temperture zero-frequency  limit of (\ref{noise def}) can be shown to be given by 
\begin{align}
\label{main result}
S({\bf r},\omega=0)&=\pi^{2}e^{4}|{\rm T}|^{4}\big[\rho_{\rm STM}(eV)\big]^{2} \chi^{\rm ch}_{\rm sub}({\bf r},\omega=0)\nonumber\\&+32\pi^{2}e^{4}|{\rm T}|^{4}\big[m_{\rm STM}(eV)\big]^{2} \chi^{\rm sp}_{\rm sub}({\bf r},\omega=0),
\end{align}
with $\rho_{\rm STM}=\sum_{\sigma}\rho^{\sigma}_{\rm STM}$, $m_{\rm STM}=\rho^{\uparrow}_{\rm STM}-\rho^{\downarrow}_{\rm STM}$, and
\begin{subequations}\label{sup}
\begin{align}
\chi^{\rm ch}_{\rm sub}({\bf r},t)&=\big<\delta \hat{n}({\bf r},t)\delta \hat{n}({\bf r},0)\big>_{H_{\rm sub}}\label{ch sup}\\
\chi^{\rm sp}_{\rm sub}({\bf r},t)&=\big<\hat{s}^{z}({\bf r},t)\hat{s}^{z}({\bf r},0)\big>_{H_{\rm sub}}\label{sp sup},
\end{align}
\end{subequations}
where $\chi^{\rm ch}_{\rm sub}$ and $\chi^{\rm sp}_{\rm sub}$ are
the local charge- and spin-susceptibilities respectively, with density and spin-density operators of the
substrate $\hat{n}$ and $\hat{s}^{z}$. The compressibility
\eqref{ch sup} is given in terms of the density variation; $\delta\hat{n}=\hat{n}-\left<\hat{n}\right>$. The extension of equation
\eqref{main result} to non-zero frequency is in principle straightforward, although relating the finite-frequency  results to physically meaningful quantities is not. This is analogous to the standard current shot-noise result, where it is only the zero-frequency component of the noise that is directly proportional to the measured current.     Equation
\eqref{main result} in
itself may not be surprising as, loosely speaking, the
conductance is determined by a single-particle correlation function
(the density of states), thus for a conductance-conductance
correlation one could expect two-particle quantities.  As a result one can obtain local susceptibilities, as a function of
position, using only a single STM or current probe.

It should also be noted that, in the weak tunneling limit the
standard expression for the current noise, which is given by the
current-current correlation function, leads to the well-known shot-noise relation.  The zero-frequency shot noise is proportional to the current
itself which goes as the tunneling amplitude squared, while
\eqref{main result} is proportional to $|{\rm T}|^{4}$. This seemingly contridictory result is explained by the fact that in general the complete characterization of
fluctuations or noise of any signal is not determined solely by a second-order
moment, such as a current-current correlation (variance), but by all higher
moments as well.  For instance one could obtain a similar result to equation \eqref{main result} from the current signal itself, by a
suitable choice of a current-current-current-current correlation (kurtosis). Also, in any real experimental measurement, the tunneling amplitude enters 
the tunneling current to all orders.  These higher order terms, the so-called vertex corrections, to the tunneling are small, and they are typically neglected; although recent STM  experiments\cite{HirjibehedinScience06,HeinrichScience04,HeinrichScience02} have shown they can lead to detectable contributions. For example, the single-triplet transition (which appears in the spin-susceptibility but not in the single-particle density of states) of an atomic spin chain has been observed.  As yet no full theoretical description of these effects exists, for one has to go beyond the standard tunneling Hamiltonian approximations\cite{PrangePR63} to describe the experimental $\partial^{2}_{V}\big<\hat{I}(eV)\big>$ curves.   Such a formulation, relating these vertex corrections to physical quantities, like spin or charge susceptibilities, is highly desirable and is of ongoing interest.  

\begin{figure}
\includegraphics[width=.45\textwidth]{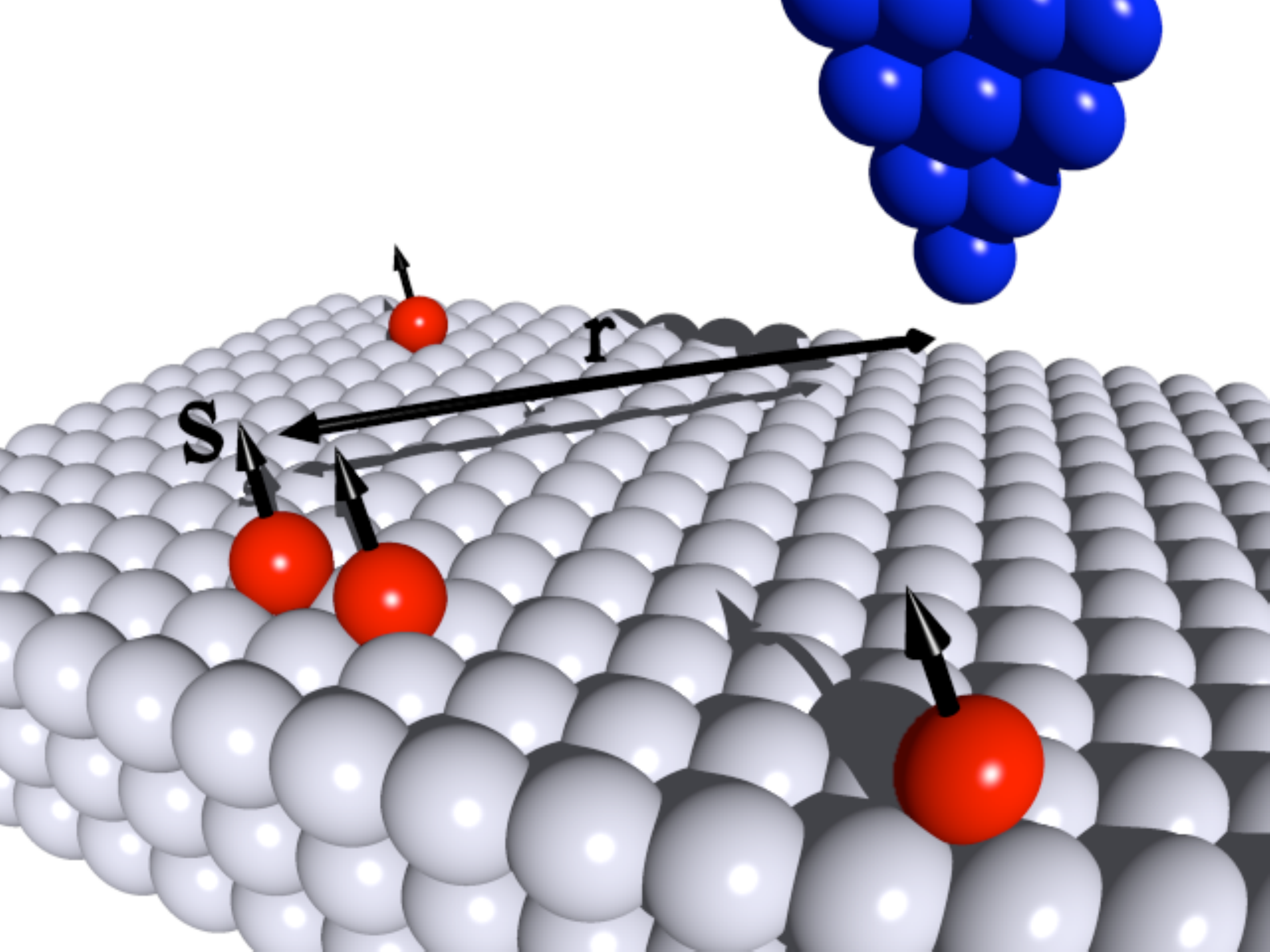}%
\caption{\label{fig1}{\bf STM setup.} Schematic of an STM and several magnetic impurities with spin ${\bf S}$ on a non-magnetic surface. Exchange between the magnetic atoms can be indirect, though the substrate, or if in close proximity, directly by the overlap of impurity wave functions. Normally, the Kondo effect is probed by tunneling into a single impurity or nano-cluster, i.e.\,${\bf r}\approx0$, where the formation of a Fano-lineshape in the current(conductance)-voltage curve is observed.}
\end{figure}

As can be seen from \eqref{main result}, for a non-magnetic STM,
i.e $m_{\rm STM}=0$, the charge susceptibility \eqref{ch sup}
determines the fluctuations, while for a spin-polarized STM
(SP-STM), $m_{\rm STM}\neq0$, the spin susceptibility \eqref{sp
sup} would be expected to dominate.  This is analogous to standard SP-STM
measurements, where the magnetic structure of the substrate is
easily resolved, even for relatively  small spin polarization
\cite{WortmannPRL01,HeinzeScience00}.  Among the many 
possibilities,  one could spatially resolve the low-energy spin
correlations near one or more Kondo impurities, where the  geometry of the nano-cluster, as well as the direct and indirect exchange processes between atoms are in strong competition\cite{SavkinPRL05} with the Kondo correlations, see Fig.~\ref{fig1}.   Here, as a simple but intriguing  example, we show that one can explore the localized spin correlations around a single magnetic atom, the so-called Kondo cloud.   We now turn to a calculation of the local
susceptibilities \eqref{ch sup} and \eqref{sp sup} for such a system.

The Kondo effect has been extensively studied both theoretically
and experimentally \cite{Hewson}, more recently by using an STM to
image single or multiple magnetic adatoms on a metallic surface
\cite{LiPRL98,MadhavenScience98,KnorrPRL02,NagaokaPRL02,WahlPRL04,NeelPRL2007}, e.g.~Fig.{\ref{fig1}}.
Experimentally, for the most part, attention has been restricted
to measuring the formation of the Abrikosov-Suhl-Kondo resonance
of  the density of states, while in the Kondo regime.
Theoretically many other quantities have been explored, such at
the non-local correlations between the impurity and conduction
electron, which is typically used to define and determine the Kondo screening
length $\xi^{}_{\rm K}$.  To calculate the required bath
correlation functions \eqref{ch sup} and \eqref{sp sup}, in the
presence of magnetic impurities, we used a numerically exact
scheme, briefly outlined below (see supplementary information for details).  In principle this formalism allows one to calculate
all physical quantities including those of the impurities, the
bath, and  all $n$-particle correlations.  Although straightforward and when applied recovers known results
\cite{SantoroPRB91b,PollweinZPhysB88, IshiiJLTP78, BordaPRB07b}, to our
knowledge it has not appeared in the literature.  We believe this is the ideal method to apply to real experimental systems.  Especially those involving  multi-impurities, in close proximity, where exchange between atoms, direct and indirect, or even the overlapping and interference of individual Kondo clouds, becomes important. 
 
The most general Hamiltonian of a noninteracting substrate with
$n$-bands coupled to $N$-atomic impurities with  amplitude
$V^{n}_{{\bf k}\sigma, \alpha s}$, and including direct exchange
$J_{\alpha, \alpha'}$ between impurities is
\begin{align}
\label{impurity ham}
&H_{\rm sub}=\sum_{n}\sum_{\bf k,\sigma}(\epsilon^{n}_{{\bf k}\sigma}-\mu)b^{\dagger}_{n,{\bf k}\sigma}b^{}_{n,{\bf k}\sigma}+\sum_{\alpha=1}^{N}\sum_{s}E_{\alpha s}c^{\dagger}_{\alpha s}c^{}_{\alpha s}\nonumber\\&+\frac{1}{2}\sum_{\alpha=1}^{N}\sum_{s_{1}\cdots s_{4}}\hspace{-1mm}U^{\alpha}_{s_{1}\cdots s_{4}}c^{\dagger}_{\alpha s_{1}}c^{\dagger}_{\alpha s_{2}}c^{}_{\alpha s_{3}}c^{}_{\alpha s_{4}}+\frac{1}{2}\hspace{-1mm}\sum_{\alpha\neq\alpha'}^{N}\hspace{-1mm}J_{\alpha,\alpha'}{\hat{\bf S}}_{\alpha}\cdot{\hat{\bf S}}_{\alpha'}\nonumber\\&+\sum_{\alpha=1}^{N}\sum_{n}\sum_{{\bf k},\sigma,s}\big[V^{n}_{{\bf k}\sigma, \alpha s}b^{\dagger}_{n,{\bf k}\sigma}c_{\alpha s}+{\rm H.c.}\big],
\end{align}
where $c^{\dagger}_{\alpha s}$ $(c^{}_{\alpha s})$ is the electron creation
(annihilation) operator for an impurity, with a complete set of
quantum numbers $s$.   Here ${\hat{\bf S}}_{\alpha}$ is the total
spin of an adatom, $E_{\alpha s}$ are the bare energy levels,
and $U^{\alpha}_{s_{1}\cdots s_{4}}$ the Coulomb interaction. In
principle all of the above parameters, along with the dispersion
of the metal $\epsilon^{n}_{{\bf k}\sigma}$, could be obtained
from an ab initio  calculation, e.g.~density functional theory.
With respect to the Hamiltonian, equation \eqref{impurity ham}, the
generating functional,  with action $S_{A},$ for the entire system
can be written as a functional integral over Grassmann variables,
including  source terms $A_{i}$ and $\bar{A}_{i}$ for the bath
electrons and each impurity; ${ Z}_{A}=\int
D[\bar{c},c]D[\bar{b},b]e^{-S_{A}}$.   Because the host metal is
assumed to be noninteracting, i.e.~Gaussian,  the bath electrons
can be integrated out exactly, leading to a reduced generating
functional ${Z}_{A}\sim\int D[\bar{c},c]e^{-S^{\rm eff}_{A}}$,
with an effective action (Hamiltonian) for the impurity sites.  The
propagator of the bath or any correlation function can
be obtain by suitable functional differentiation of the effective
action with respect to the sources.  Doing so, the only
unknown correlators are those of the impurities. The evaluation of which can be done using a variety of computationally fast and accurate impurity
solvers.  Here, we used the numerically exact
continuous-time quantum Monte Carlo (CT-QMC) method of
Ref.~[\onlinecite{RubtsovPRB05}].
\begin{figure}
\includegraphics[width=.48\textwidth]{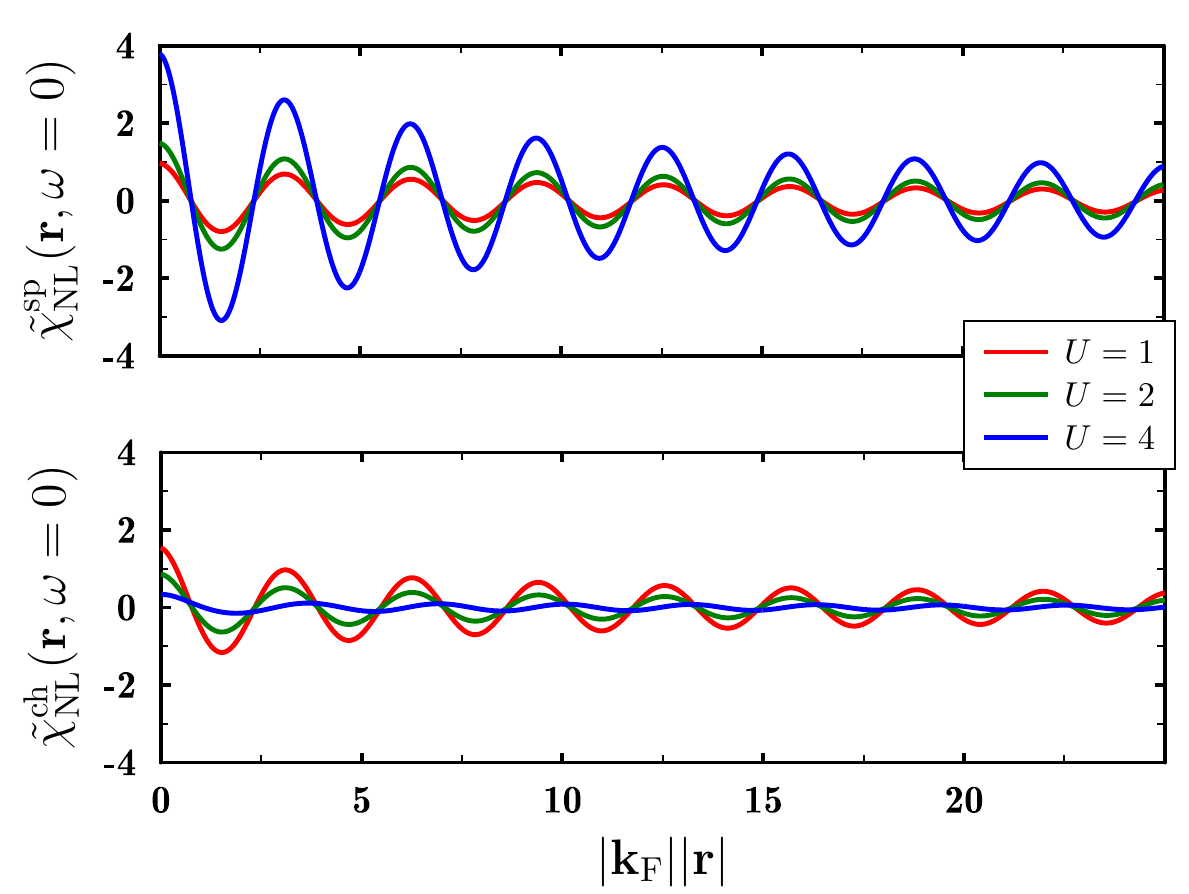}%
\caption{{\bf Non-local impurity-bath correlations.} The zero-frequency  non-local (NL) spin (top) and charge (bottom) correlations, defined as  $\chi^{\rm sp}_{\rm NL}({\bf r},t)=\big<\hat{s}^{z}({\bf r},t)\hat{S}^{z}_{\rm imp}(0)\big>$ and $\chi^{\rm ch}_{\rm NL}({\bf r},t)=\big<\delta\hat{n}({\bf r},t)\delta\hat{N}_{\rm imp}(0)\big>$, between a magnetic impurity and the conduction electrons as a function of distance from the impurity. In the graph $\tilde{\chi}^{}_{\rm NL}=\frac{2|{\bf k}_{\rm F}|^{2}|{\bf r}|^{2}}{\pi N_{\sigma}\Gamma}\chi^{}_{\rm NL}$.  All other relevant parameters  are given in Table \ref{table1}.\label{fig2}}
\end{figure}

\begin{table}
\begin{center}
\begin{tabular}{cclcc|c|cc}
\toprule
$\epsilon_{\rm F}$&&$E^{}_{c}$& &$U$&$T^{}_{\rm K}$&$\xi^{}_{\rm K}$&\\
\hline
6.0&&-0.5& & 1.0 & 0.168&$\sim$ 70&\\
\hline
6.0&&-1.0& & 2.0 & 0.0865&$\sim$ 140&\\
\hline
6.0&&-2.0& & 4.0 & 0.0216&$\sim$ 555&\\
\hline
\end{tabular}
\caption{Parameters for the symmetric single-impurity spin-$1/2$ Anderson model in energy units of $\Gamma=\pi |V|^{2}N_{\sigma}({\epsilon}_{\rm F})$, where $\epsilon_{\rm F}$ and $N_{\sigma}({\epsilon}_{\rm F})$ are the Fermi energy and spin-resolved noninteracting density of states of the bath respectively. Lengths are in units of the inverse Fermi wavevector $|{\bf k}_{\rm F}|^{-1}$.  The Kondo temperature $T_{\rm K}$ is obtained from the Bethe-Ansatz solution \cite{WiegmannJPhysC83a}. The expected size of the Kondo screening  cloud  is of order $\xi^{}_{\rm K}\approx \hbar v_{\rm F}/(k^{}_{\rm B}T^{}_{\rm K})$, where $v_{\rm F}$ is the Fermi velocity. Unless otherwise stated all calculations were done at an inverse-temperature of $(k^{}_{\rm B}T)^{-1}=\beta=200$, well below the Kondo temperature for these parameters.}\label{table1}
\end{center}
\end{table}

As an example of the usefulness and flexibility of the above
formalism, we calculated the zero-frequency non-local charge and
spin correlations between an impurity and the bath
(Fig.~\ref{fig2}), which has been extensively studied both
analytically and numerically for equal times\cite{IshiiJLTP78,PollweinZPhysB88,GubernatisPRB87,SorensenPRB96,BordaPRB07}, i.e.~$t=0$. 
For simplicity and clarity, the real Hamiltonian \eqref{impurity ham} is
approximated by the symmetric single-impurity spin-$1/2$ Anderson
model \cite{Hewson}, with an onsite $U$ and a 3D parabolic
dispersion for the bath. We will also neglect the direct tunneling into the impurity.  With this contribution our results would be modified only for ${\bf r}\approx 0$.   In the low-energy or low-frequency
regime, Fig.~\ref{fig2} clearly shows a separation of scales
between spin and charge correlations as the on-site Coulomb energy
$U$ is increased, as compared to the case of equal times
(effectively high-energy),
e.g.~Ref.~[\onlinecite{GubernatisPRB87}].  This is what one would
expect as  the dominate correlations of the Kondo effect are at low
energy, while charge fluctuations of the impurity are suppressed with increasing
$U$.

\begin{figure}
\includegraphics[width=.48\textwidth]{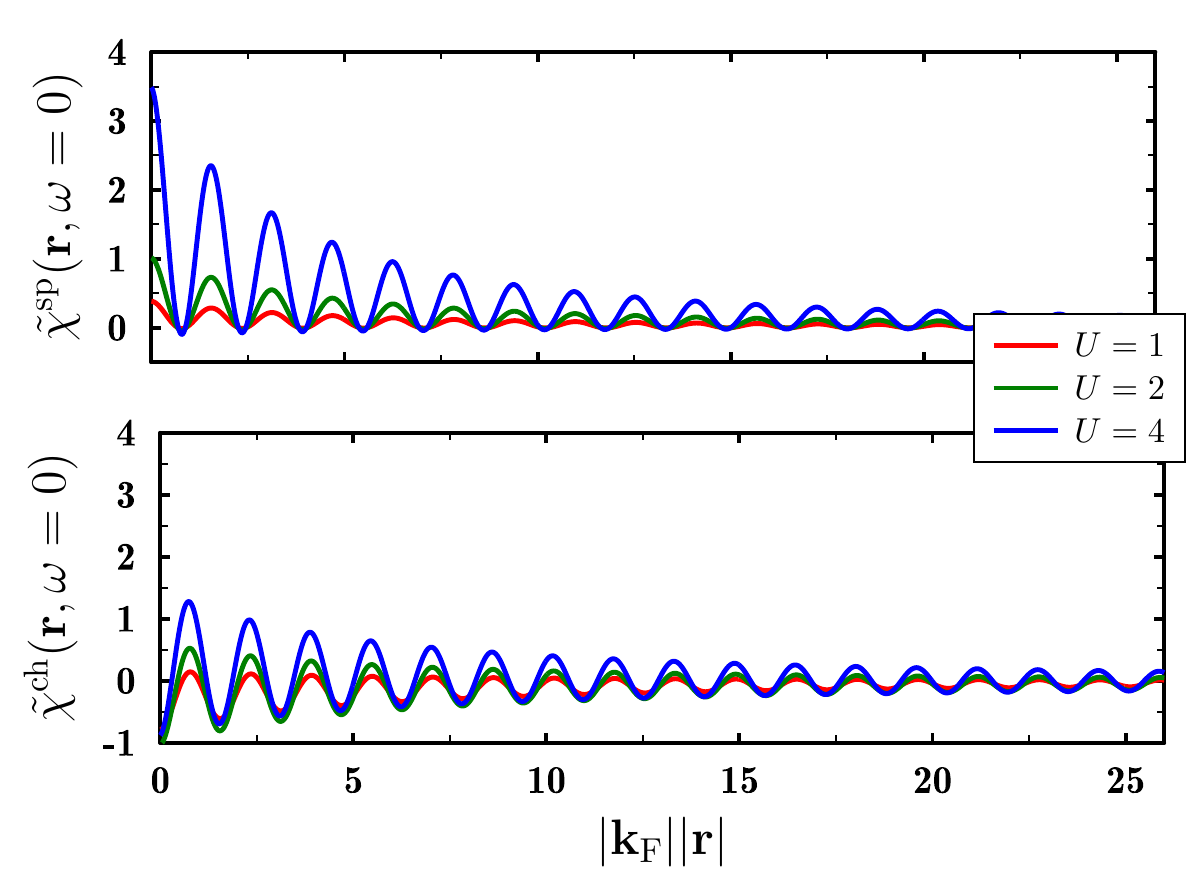}%
\caption{{\bf Local correlations of the bath.} The zero-frequency local spin (top) and charge (bottom) susceptibilities, equations \eqref{ch sup} and \eqref{sp sup}, of the bath for the single-impurity spin-$1/2$ Anderson model as a function of the distance from the impurity site. Here, $\tilde{\chi}^{}_{}=\big[\frac{2\Gamma |{\bf k}_{\rm F}|  |{\bf r}|}{\pi N_{\sigma}}\big]^{2}\chi^{}_{}$.   All parameters are the same as in Fig.~\ref{fig2} and are given in Table \ref{table1}. \label{fig3}}
\end{figure}

\begin{figure}
\includegraphics[width=.5\textwidth]{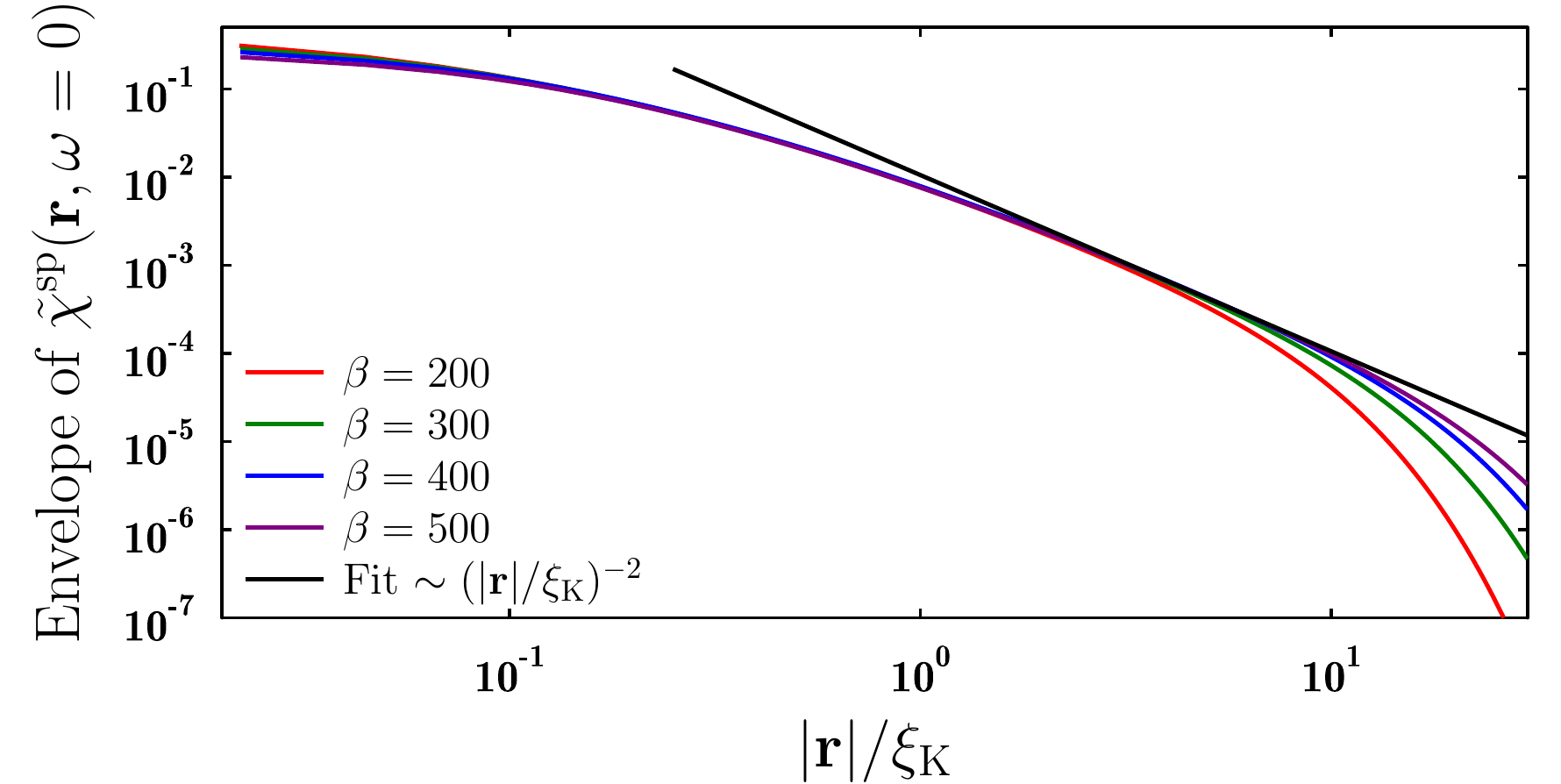}%
\caption{{\bf Envelope of spin-spin correlation.} The envelope of the zero-frequency local  spin susceptibility  (equation \eqref{sp sup}, shown in Fig.~\ref{fig2}) for fixed $U=1$ and different temperatures $T$; $\beta=(k^{}_{\rm B}T)^{-1}$.  The horizontal axis has been rescaled by the Kondo length; $\xi^{}_{\rm K}= \hbar v_{\rm F}/(k^{}_{\rm B}T^{}_{\rm K})$. All other parameters are given in Table \ref{table1}.   \label{fig4}}
\end{figure}
In Fig.~\ref{fig3} the zero-frequency local charge and spin
susceptibilities of the bath  [equations \eqref{ch sup} and \eqref{sp
sup}], which determine the conduction fluctuations, for different
$U$ are shown.  The oscillation period of both is $\pi/2$, the
same as Friedel oscillations, but the spin correlation function is
$\pi/2$ phase shifted from that of the charge.  The envelope or
decay of the spin correlations as a function of distance is shown
in Fig.~\ref{fig4}. For ${\bf r}\ll\xi^{}_{\rm K}$  the
correlations  show a non-algebraic decay, which changes to a
power-law for ${\bf r}\approx \xi^{}_{\rm K}$, and ultimately, at
finite temperature, the correlations are exponentially cut-off by
the thermal length $\xi^{}_{T}\sim \hbar v^{}_{\rm F}/k^{}_{\rm
B}T$.  The appearance of the thermal length can be seen for the
largest values of ${\bf r}$ in Fig.~\ref{fig4}, as the power-law
changes over into an exponential.  At zero-temperature the decay would
remain a power-law for ${\bf r}\gg \xi^{}_{\rm K}$.  The simple physical interpretation of this is:
at zero-temperature the magnetic impurity is almost fully screened by
conduction electrons within $\xi^{}_{\rm K}$, thus outside of this
length scale correlations are weak and decay quite rapidly, but
within $\xi^{}_{\rm K}$ correlations of the bath, mediated by the
impurity,  remain non-trivial.

In conclusion we have shown that the conductance fluctuations of a
tunneling current into an interacting system is determined by the
charge and spin susceptibilities of the system.  We have also shown that one application of this is to use an SP-STM to
detect the Kondo screening length, $\xi^{}_{\rm K}$, for a
magnetic adatom on a metallic surface. Although the Kondo problem has been and continues to be one of the most intensively studied phenomena in condensed matter physics,  there has yet to be an experimental conformation of this theoretical prediction concerning  the screening cloud.  We have furthermore developed a general method to exactly calculate $n$-point correlations for experimentally relevant setups consisting of multiple adatoms or correlated ``sites''.  
 Extension of these results to superconducting systems and quantum dot geometries  is of future interest.

\begin{acknowledgments}
This work has been supported by  the
German Research Council (DFG) under SFB 668. K.P. would like thank Kirsten von Bergmann, Germar Hoffmann, V\'aclav Jani\v{s}, Hartmut Monien, and Jens Wiebe for useful discussions.
\end{acknowledgments}

\onecolumngrid
\renewcommand{\thesection}{S\Roman{section}}
\renewcommand{\theequation}{S\Roman{section}-\arabic{equation}}
\section*{SUPPLEMENTARY INFORMATION}
\setcounter{equation}{0}
\numberwithin{equation}{section}
\section{Conductance operator and spectral density}
Quantum mechanically the fluctuations, or uncertainty, of an observable is related to the variance of its expectation value.  To obtain an expression for the conductance fluctuations or more specifically the spectral density, one needs an operator for the conductance. This  can be found by taking 
the derivative of the linear response expression of the tunneling current, equation (\ref{LR current operator}), with respect to the
applied voltage, $V$. Doing so gives 
\begin{align}
\label{LR conductance operator}
\hat{G}^{}_{\rm LR}({\bf r},t)&=ie^{2}|{\rm T}|^{2}\sum_{\sigma,\sigma'}\int\limits_{-\infty}^{\infty}dt'\, \Theta(t-t')(t-t')\Big\{\big[C^{}_{\sigma'}({\bf r},t'),C^{\dagger}_{\sigma}({\bf r},t)\big]-\big[C^{}_{\sigma}({\bf r},t),C^{\dagger}_{\sigma'}({\bf r},t')\big]\Big\}\nonumber\\&-ie^{2}|{\rm T}|^{2}\sum_{\sigma,\sigma'}\int\limits_{-\infty}^{\infty}dt'\, \Theta(t-t')(t+t')\Big\{\big[C^{}_{\sigma}({\bf r},t),C^{}_{\sigma'}({\bf r},t')\big]+\big[C^{\dagger}_{\sigma}({\bf r},t),C^{\dagger}_{\sigma'}({\bf r},t')\big]\Big\},
\end{align}
where $C_{\sigma}({\bf r})$ is defined in terms of the field
operators of the STM and substrate;  $C_{\sigma}({\bf
r})=\Psi^{\dagger}_{\rm STM}({\bf r}\sigma)\Psi^{}_{\rm sub}({\bf
r}\sigma)$. Although the tunneling matrix elements, $\rm T$, in \eqref{LR conductance operator} have different dimensions than that of \eqref{full current operator}, we will use the same symbol for simplicity.  Upon taking the expectation value, the second term of
(\ref{LR conductance operator}) generates anomalous Green's
functions that only contribute for a supercurrent. These terms are
responsible for the Josephson effect and  will not be considered
here.    For a non-superconducting system the expectation of
(\ref{LR conductance operator}) recovers the well-known expression
for the conductance in terms of the local single-particle density
of states $\rho^{\sigma}({\bf r},\omega)$; $\big<\hat{G}^{}_{\rm
LR}\big>_{H^{}_{0}}=2\pi e^{2}|{\rm
T}|^{2}\sum_{\sigma}\rho^{\sigma}_{\rm STM}(\epsilon^{}_{\rm F})\rho^{\sigma}_{\rm
sub}({\bf r},eV)$. 

The symmetrized spectral density that characterizes the frequency distribution of fluctuations of the conductance about its averaged value is defined as
\begin{align}
\label{noise def in supp}
S({\bf r},\omega)&=\frac{1}{2}\int dt\, e^{i\omega t}\big<\big\{\delta \hat{G}_{\rm LR}({\bf r},t),\delta \hat{G}_{\rm LR}({\bf r},0)\big\}\big>_{H_{0}},
\end{align}
where $\delta \hat{G}_{\rm LR}=\hat{G}_{\rm LR}-\big<\hat{G}_{\rm LR}\big>$.  This expression for the spectral density assumes time-translational invariance, although the full expression for the conductance operator (\ref{LR conductance operator}) contains non time-translational invariant terms, these vanish for non-superconducting systems. Using \eqref{LR conductance operator} and \eqref{noise def in supp}, a lengthy, but straightforward, calculation (\ref{noise def in supp}) gives the zero-frequency component of the noise as
\begin{align}
\label{main result in supp}
S({\bf r},\omega=0)&=8\pi^{2}e^{4}|{\rm T}|^{4}\big[\rho_{\rm STM}(eV)\big]^{2} \chi^{\rm ch}_{\rm sub}({\bf r},\omega=0)+32\pi^{2}e^{4}|{\rm T}|^{4}\big[m_{\rm STM}(eV)\big]^{2} \chi^{\rm sp}_{\rm sub}({\bf r},\omega=0). 
\end{align}
Equation \eqref{main result in supp} is the starting point and motive for the remaining  calculations and results of the article. 
Technically, only the vertex corrections of \eqref{ch
sup} and \eqref{sp sup} are measured in \eqref{main result in supp}, but in the
low-temperature limit, the single-particle contribution to these
correlation functions is negligible and vanishes at zero temperature. Thus, we will work with the full correlation
functions for simplicity and clarity. 

\section{Generating functional for N-correlated sites}
In this section we show how one can, in a simple and straightforward manner, obtain arbitrary correlation functions of a system consisting of $N$-correlated sites in contact with a non-interacting bath.  In essence this method reduces {\it all} correlations to correlations involving only the impurity operators.  The remaining impurity problem can then be solved using a variety of techniques, such as numerical renormalization  group or quantum Monte Carlo.  Theoretically the number of impurities $N$ can be arbitrary, but in practice it is limited by the complexity of the system at hand, such as the number of bands of the substraight and orbitals of the adatoms.  In the simplest case, for two-particle properties, an upper bound of four or five Anderson impurities could be incorporated  using the CT-QMC impurity solver used in this work.  

The partition function $Z={\rm Tr}\,e^{-\beta H}$ with respect to the $N$-impurity Hamiltonian \eqref{impurity ham}  can be written as a functional integral over Grassmann fields (by inserting  complete sets of fermionic coherent states) as
\begin{equation}
Z=\prod_{\alpha=1}^{N}\int { D}[\bar{b},{b}] {D}[\bar{c}^{}_{\alpha},{c}^{}_{\alpha}]e^{-S}
\end{equation}
with an action $S$ in Fourier-space 
\begin{align}
S&=\sum_{n}\sum_{{\bf k}\sigma}\sum_{i\omega^{}_{m}}\bar{b}^{n}_{{\bf k}\sigma}(i\omega^{}_{m})\big[i\omega^{}_{m}-(\epsilon^{n}_{{\bf k}\sigma}-\mu)\big]{b}^{n}_{{\bf k}\sigma}(i\omega^{}_{m})+\sum_{\alpha,s}\sum_{i\omega^{}_{m}}\bar{c}^{}_{\alpha s}(i\omega^{}_{m})(i\omega^{}_{m}-E^{}_{\alpha s}){c}^{}_{\alpha s}(i\omega^{}_{m})\nonumber\\&+\sum_{n, \alpha, s}\sum_{{\bf k}\sigma }\sum_{i\omega^{}_{m}}\big[V^{n}_{{\bf k}\sigma,\alpha s}\bar{b}^{n}_{{\bf k}\sigma}(i\omega^{}_{m}){c}^{}_{\alpha s}(i\omega^{}_{m})+V^{*n}_{{\bf k}\sigma,\alpha s}\bar{c}^{}_{\alpha s}(i\omega^{}_{m}){b}^{n}_{{\bf k}\sigma}(i\omega^{}_{m})\big]+S_{\rm int}[\bar{c}^{}_{\alpha s}(i\omega^{}_{m}),{c}^{}_{\alpha s}(i\omega^{}_{m})],
\end{align}
where $i\omega^{}_{m}$ are fermionic Matsubara frequencies and $S_{\rm int}$ is the term of the action (Hamiltonian) that corresponds to the interactions of the $N$ impurities.  The actual form of which is immaterial, as long as it is a local interaction.  
By introducing Grassmann source terms, $A$ and $\bar{A}$ for each impurity and the bath electrons, the generating functional $Z_{A}$ for the system is given by 
\begin{align}
\label{generating functional}
Z_{A}=\prod_{\alpha=1}^{N}\int { D}[\bar{b},{b}] {D}[\bar{c}^{}_{\alpha},{c}^{}_{\alpha}]e^{-S}\exp&\bigg\{\sum_{\alpha,s}\sum_{i\omega^{}_{m}}\big[\bar{A}_{\alpha s}(i\omega_{m}){c}^{}_{\alpha s}(i\omega^{}_{m})+{A}_{\alpha s}(i\omega_{m})\bar{c}^{}_{\alpha s}(i\omega^{}_{m})\big]\nonumber\\ &+\sum_{n}\sum_{{\bf k}\sigma}\sum_{i\omega^{}_{m}}\big[\bar{A}^{n}_{{\bf k}\sigma}(i\omega_{m}){b}^{n}_{{\bf k}\sigma}(i\omega^{}_{m})+{A}^{n}_{{\bf k}\sigma}(i\omega_{m})\bar{b}^{n}_{{\bf k}\sigma}(i\omega^{}_{m})\big]\bigg\}. 
\end{align}
Because we have assumed a non-interacting bath and local interactions of the impurities, the  bath electrons remain Gaussian and with the identity

\begin{align}
\label{integration identity}
&\int D[\bar{\eta},\eta]\exp\Big[{-\iint dxdx'\bar{\eta}(x)M(x,x')\eta(x')+\int dx J^{}(x)\bar{\eta}(x)}{\xi}(x)+\int dx J^{*}(x)\bar{\xi}(x)\eta(x)\Big]\nonumber\\&=\det[M(x,x')]\exp\Big[\iint dx dx'J^{*}(x){\bar\xi}(x)M^{-1}(x,x'){\xi}(x')J^{}(x')\Big],
\end{align}
can be exactly integrated out of  the generating functional \eqref{generating functional}, leading to 
\begin{align}
\label{reduced generating functional}
&Z_{A}=\det\big[i\omega^{}_{m}-(\epsilon^{n}_{{\bf k}\sigma}-\mu)\big]\prod_{\alpha=1}^{N}\int  {D}[\bar{c}^{}_{\alpha},{c}^{}_{\alpha}]e^{-S_{\rm eff}} \exp\Bigg\{\sum_{\alpha,s}\sum_{i\omega^{}_{m}}\big[\bar{A}_{\alpha s}(i\omega_{m}){c}^{}_{\alpha s}(i\omega^{}_{m})+{A}_{\alpha s}(i\omega_{m})\bar{c}^{}_{\alpha s}(i\omega^{}_{m})\big]\nonumber\\&+\sum_{n,\alpha,s}\sum_{{\bf k}\sigma}\sum_{i\omega^{}_{m}} \frac{V^{n}_{{\bf k}\sigma,\alpha s}{c}^{}_{\alpha s}(i\omega^{}_{m})\bar{A}^{n}_{{\bf k}\sigma}(i\omega^{}_{m})}{i\omega^{}_{m}-(\epsilon^{n}_{{\bf k}\sigma}-\mu)}+\sum_{n,\alpha,s}\sum_{{\bf k}\sigma}\sum_{i\omega^{}_{m}} \frac{{A}^{n}_{{\bf k}\sigma}(i\omega^{}_{m})V^{*n}_{{\bf k}\sigma,\alpha s}\bar{c}^{}_{\alpha s}(i\omega^{}_{m})}{i\omega^{}_{m}-(\epsilon^{n}_{{\bf k}\sigma}-\mu)}+\sum_{n}\sum_{{\bf k}\sigma}\sum_{i\omega^{}_{m}}\frac{A^{n}_{{\bf k}\sigma}(i\omega^{}_{m})\bar{A}^{n}_{{\bf k}\sigma}(i\omega^{}_{m})}{i\omega^{}_{m}-(\epsilon^{n}_{{\bf k}\sigma}-\mu)}\Bigg\},
\end{align}
where the effective action of the impurities is, in matrix notation, 
\begin{equation}
S_{\rm eff}=-\sum_{s,s'}\sum_{i\omega_{m}}{\bf \bar{c}}^{T}_{s}(i\omega_{m}){\bf {\cal G}_{0}^{-1}}(i\omega_{m}){\bf{c}}^{}_{s'}(i\omega_{m})+S_{\rm int}[\bar{c}^{}_{\alpha s}(i\omega^{}_{m}),{c}^{}_{\alpha s}(i\omega^{}_{m})]
\end{equation}
with
\begin{equation*}
{\bf \bar{c}}^{}_{s}(i\omega_{m})=\left(\begin{array}{c}\bar{c}_{s\alpha=1}(i\omega^{}_{m}) \\\vdots\\ \bar{c}_{s\alpha=N}(i\omega^{}_{m})\end{array}\right),\hspace{1cm}{\bf {c}}^{}_{s}(i\omega_{m})=\left(\begin{array}{c}{c}_{s\alpha=1}(i\omega^{}_{m}) \\\vdots\\ {c}_{s\alpha=N}(i\omega^{}_{m})\end{array}\right),
\end{equation*}
${\bf \bar c}^{T}$ being the transpose of the column vector, and 
\begin{equation*}
\big[{\bf {\cal G}_{0}^{-1}}(i\omega_{m})\big]_{\alpha s, \alpha' s'}=(i\omega^{}_{m}-E_{\alpha s })\delta_{\alpha,\alpha'}\delta_{s,s'} -\Delta_{\alpha s,\alpha' s'}(i\omega^{}_{m}),
\end{equation*}
where the hybridization or Weiss field is
\begin{equation*}
\Delta_{\alpha s ,\alpha' s'}(i\omega^{}_{m})=\sum_{n,{\bf k}\sigma}\frac{V^{n}_{{\bf k}\sigma,\alpha s}V^{*n}_{{\bf k}\sigma,\alpha' s'}}{i\omega^{}_{m}-(\epsilon^{n}_{{\bf k}\sigma}-\mu)}. 
\end{equation*}
The determinant that appears in \eqref{reduced generating functional} can be analytically evaluated, but it is actually never referenced in a calculation of a correlation function, as it is canceled by an identical term appearing in an overall normalization factor. 

Using \eqref{reduced generating functional}, arbitrary correlation functions can be obtained by suitable functional differention.  As can be seen, bath correlators (apart from trivial non-interacting terms) are found from  impurity correlators simply by attaching ``tails'' of the form $\frac{V^{n}_{{\bf k}\sigma,\alpha s}}{i\omega^{}_{m}-(\epsilon^{n}_{{\bf k}\sigma}-\mu)}$ or $\frac{V^{*n}_{{\bf k}\sigma,\alpha s}}{i\omega^{}_{m}-(\epsilon^{n}_{{\bf k}\sigma}-\mu)}$.       For example the single-particle Green's function of the bath is
\begin{align}
&G^{n,n'}_{{\bf k}\sigma,{\bf k}'\sigma'}(i\omega^{}_{m},i\omega^{}_{m'})=\frac{1}{Z^{}_{A}\big|_{A=0}}\frac{\delta^{2}Z^{}_{A}}{\delta A^{n}_{{\bf k}\sigma}(i\omega^{}_{m})\delta \bar{A}^{n'}_{{\bf k'}\sigma'}(i\omega^{}_{m'})}\Bigg|_{A=0}\nonumber\\&=\frac{1}{i\omega^{}_{m}-(\epsilon^{n}_{{\bf k}\sigma}-\mu)}\delta_{{\bf k},{\bf k'}}\delta_{\sigma,\sigma'}\delta_{i\omega^{}_{m},i\omega^{}_{m'}}\delta_{n,n'}-\sum_{\alpha,\alpha'}\sum_{s,s'}\frac{V^{n}_{{\bf k}\sigma,\alpha s}}{i\omega^{}_{m}-(\epsilon^{n}_{{\bf k}\sigma}-\mu)}\big<c^{}_{\alpha,s}(i\omega^{}_{m})\bar{c}^{}_{\alpha',s'}(i\omega^{}_{m'})\big>\frac{V^{*n'}_{{\bf k}'\sigma',\alpha' s'}}{i\omega^{}_{m'}-(\epsilon^{n'}_{{\bf k}'\sigma'}-\mu)},
\end{align}
which is just the multi-impurity generalization of the well-known $T$-matrix expression, commonly found by equation of motion methods.  Similarly the two-particle Green's function of the bath is given by
\begin{align}
&G^{n_{1},n_{2},n_{3},n_{4}}_{{\bf k}_{1}\sigma_{1},{\bf k}_{2}\sigma_{2},{\bf k}_{3}\sigma_{3},{\bf k}_{4}\sigma_{4}}(i\omega^{}_{1},i\omega^{}_{2},i\omega^{}_{3},i\omega^{}_{4})=\nonumber\\&G^{n_{1},n_{2}}_{{\bf k}_{1}\sigma_{1},{\bf k}_{2}\sigma_{2}}(i\omega^{}_{1},i\omega^{}_{2})G^{n_{3},n_{4}}_{{\bf k}_{3}\sigma_{3},{\bf k}_{4}\sigma_{4}}(i\omega^{}_{3},i\omega^{}_{4})-G^{n_{1},n_{4}}_{{\bf k}_{1}\sigma_{1},{\bf k}_{4}\sigma_{4}}(i\omega^{}_{1},i\omega^{}_{4})G^{n_{3},n_{2}}_{{\bf k}_{3}\sigma_{3},{\bf k}_{2}\sigma_{2}}(i\omega^{}_{3},i\omega^{}_{2})\nonumber\\&+\sum_{\stackrel{\alpha_{1},\alpha_{2},\alpha_{3},\alpha_{4}}{s_{1},s_{2},s_{3},s_{4}}}\frac{V^{n_{1}}_{{\bf k}_{1}\sigma_{1},\alpha_{1} s_{1}}}{i\omega^{}_{1}-(\epsilon^{n_{1}}_{{\bf k}_{1}\sigma_{1}}-\mu)}\frac{V^{*n_{2}}_{{\bf k}_{2}\sigma_{2},\alpha_{2} s_{2}}}{i\omega^{}_{2}-(\epsilon^{n_{2}}_{{\bf k}_{2}\sigma_{2}}-\mu)}{\Delta}^{\alpha_{1},\alpha_{2},\alpha_{3},\alpha_{4}}_{s_{1},s_{2},s_{3},s_{4}}(i\omega^{}_{1},i\omega^{}_{2},i\omega^{}_{3},i\omega^{}_{4})\frac{V^{n_{3}}_{{\bf k}_{3}\sigma_{3},\alpha_{3} s_{3}}}{i\omega^{}_{3}-(\epsilon^{n_{3}}_{{\bf k}_{3}\sigma_{3}}-\mu)}\frac{V^{*n_{4}}_{{\bf k}_{4}\sigma_{4},\alpha_{4} s_{4}}}{i\omega^{}_{4}-(\epsilon^{n_{4}}_{{\bf k}_{4}\sigma_{4}}-\mu)},
\end{align}
where
\begin{equation}
\Delta^{\alpha_{1},\alpha_{2},\alpha_{3},\alpha_{4}}_{s_{1},s_{2},s_{3},s_{4}}(i\omega^{}_{1},i\omega^{}_{2},i\omega^{}_{3},i\omega^{}_{4})=\sum_{\stackrel{\alpha'_{1},\alpha'_{2},\alpha'_{3},\alpha'_{4}}{s'_{1},s'_{2},s'_{3},s'_{4}}}G^{\alpha_{1},\alpha'_{1}}_{s_{1},s'_{1}}(i\omega^{}_{1})G^{\alpha_{2},\alpha'_{2}}_{s_{2},s'_{2}}(i\omega^{}_{2}){\Gamma}^{\alpha'_{1},\alpha'_{2},\alpha'_{3},\alpha'_{4}}_{s'_{1},s'_{2},s'_{3},s'_{4}}(i\omega^{}_{1},i\omega^{}_{2},i\omega^{}_{3},i\omega^{}_{4})G^{\alpha'_{3},\alpha_{3}}_{s'_{3},s_{3}}(i\omega^{}_{3})G^{\alpha'_{4},\alpha_{4}}_{s'_{4},s_{4}}(i\omega^{}_{4})
\end{equation} 
 is the non-trival part of the impurity two-particle Green's function,  i.e.,
\begin{align}
&G^{\alpha_{1},\alpha_{2},\alpha_{3},\alpha_{4}}_{s_{1},s_{2},s_{3},s_{4}}(i\omega^{}_{1},i\omega^{}_{2},i\omega^{}_{3},i\omega^{}_{4})=G^{\alpha_{1},\alpha_{2}}_{s_{1},s_{2}}(i\omega^{}_{1},i\omega^{}_{2})G^{\alpha_{3},\alpha_{4}}_{s_{3},s_{4}}(i\omega^{}_{3},i\omega^{}_{4})-G^{\alpha_{1},\alpha_{4}}_{s_{1},s_{4}}(i\omega^{}_{1},i\omega^{}_{4})G^{\alpha_{3},\alpha_{2}}_{s_{3},s_{2}}(i\omega^{}_{3},i\omega^{}_{2})\nonumber\\&+\sum_{\stackrel{\alpha'_{1},\alpha'_{2},\alpha'_{3},\alpha'_{4}}{s'_{1},s'_{2},s'_{3},s'_{4}}}G^{\alpha_{1},\alpha'_{1}}_{s_{1},s'_{1}}(i\omega^{}_{1})G^{\alpha_{2},\alpha'_{2}}_{s_{2},s'_{2}}(i\omega^{}_{2}){\Gamma}^{\alpha'_{1},\alpha'_{2},\alpha'_{3},\alpha'_{4}}_{s'_{1},s'_{2},s'_{3},s'_{4}}(i\omega^{}_{1},i\omega^{}_{2},i\omega^{}_{3},i\omega^{}_{4})G^{\alpha'_{3},\alpha_{3}}_{s'_{3},s_{3}}(i\omega^{}_{3})G^{\alpha'_{4},\alpha_{4}}_{s'_{4},s_{4}}(i\omega^{}_{4}),
\end{align}
here $\Gamma$ is the reducible vertex of the impurity problem. 
\end{document}